\documentclass[11pt,a4paper]{article}
\usepackage{graphicx}
\usepackage[unicode=true]{hyperref}
\usepackage{float}
\usepackage{amsmath, amssymb, amsthm, accents}
\usepackage{centernot}
\usepackage{bm}
\usepackage{booktabs}
\usepackage{natbib}
\setcitestyle{authoryear,round,semicolon}

\usepackage{enumitem}
\setlist[itemize]{itemsep=1pt} 
\setlist[enumerate]{itemsep=1pt} 

\usepackage{caption}
\usepackage{subcaption} 

\makeatletter
\def\widebar{\accentset{{\cc@style\underline{\mskip10mu}}}}
\makeatother

\title{Uncovering Functional Blocks in Interregional Production Networks:
Evidence from Input–Output Linkages in Japan\footnote{I am grateful to participants at the 5th Annual Meeting of the Computational Social Science Society of Japan (CSSJ 2026) for valuable comments. This work was supported by JSPS KAKENHI Grant Number 22K18525. All remaining errors are my own.}}
\author{Shota Fujishima\footnote{Graduate School of Economics, Hitotsubashi University, 2-1 Naka, Kunitachi, Tokyo 186-8601, Japan. Email: {\tt s.fujishima@r.hit-u.ac.jp}; Website: \texttt{https://sites.google.com/view/sfujishima}}}
\date{\today}

\begin{document}

\maketitle

\begin{abstract}
This paper examines the latent functional block structure of Japan’s production network using interregional input–output data. To isolate non-trivial production linkages, we first estimate a structural gravity model to account for spatial frictions and economic scale, and then apply a weighted stochastic blockmodel (SBM) to the resulting residual network. Because these residual linkages often connect distant regions, the SBM is well suited to grouping region–industry pairs based on their shared macroeconomic roles. The results reveal that even after explicitly filtering out the mechanical effects of geographic proximity, the network is organized into functional blocks that maintain a high degree of regional coherence. Beyond this baseline spatial clustering, we find evidence of cross-regional integration, a structural bifurcation between manufacturing and urban services in metropolitan areas, and broadly spanning primary sectors. These findings provide a network-based perspective on regional coordination, offering guidance for how structurally defined production blocks—rather than simple geographic proximity—can inform wide-area policy design.

\end{abstract}

\section{Introduction}
In the context of population decline and rapid population aging, the importance
of inter-municipal cooperation has become increasingly recognized in regional
development policies in Japan.
One prominent example is the Network-based Core City Region Initiative promoted
by the Ministry of Internal Affairs and Communications.
According to the official guideline,\footnote{\href{https://www.soumu.go.jp/main_sosiki/jichi_gyousei/renkeichusutoshiken/index.html}{https://www.soumu.go.jp/main\_sosiki/jichi\_gyousei/renkeichusutoshiken/index.html (in Japanese)}}
the initiative aims to form regional hubs that maintain economic vitality even
under demographic decline, by encouraging a central city with sufficient scale
and urban functions to cooperate with its surrounding municipalities.
Through compact urban development and network-based coordination, the policy
seeks to pool resources and leverage respective comparative advantages across
administrative boundaries.

A key implicit assumption behind such policies is that effective economic
cooperation is largely dictated by geographic proximity.
Spatial closeness undoubtedly facilitates administrative coordination and
daily-life services.
However, modern economies are characterized by nationwide and highly
intricate webs of industrial interdependencies that extend far beyond local
administrative borders.
Understanding how regional economies are embedded in this broader production
network is therefore crucial for designing effective regional hubs.
In particular, an important empirical question arises: do local economies cluster primarily based on geographic proximity, or do they group into broader structural blocks defined by shared functional roles within the national supply chain?

Recent advances in network science highlight an important methodological
challenge when studying spatial interaction networks.
Because geographic distance strongly shapes link formation, conventional
community detection algorithms often recover geographically contiguous
clusters even when spatial proximity alone explains most interactions.
For instance, \citet{expert2011ProcNatlAcadSciUA} show that applying standard community
detection methods to spatial networks tends to identify communities that
merely reflect distance-driven connectivity patterns.
To address this issue, they propose spatial null models that explicitly
account for geographic distance.
Similar concerns arise in economic networks such as international trade,
where gravity forces driven by economic size and distance dominate
interaction patterns.
A growing body of research therefore constructs gravity-based benchmark
networks and examines the residual structures that remain after controlling
for these pervasive forces \citep[e.g.,][]{fagiolo2010JEconInteractCoord, duenas2013JEconInteractCoord, mastrandrea2025SocialNetworks}.
These studies suggest that economically meaningful structures are often
revealed not in the raw interaction network itself, but in the deviations
from gravity-like baselines.

Building on this insight, this study aims to uncover the latent architecture
of interregional production networks in Japan.
Specifically, we analyze interactions among ``region--industry'' pairs
(e.g., agriculture in Hokkaido or construction in the Kanto region),
which represent economically meaningful units for regional policy analysis.
To systematically identify this architecture, we propose a two-step empirical
framework using a Japanese interregional input--output table.
First, drawing on the structural gravity literature extensively developed
in international trade \citep{anderson2003Am.Econ.Rev.}, we estimate a gravity
model to filter out the pervasive effects of economic size and geographic
distance on trade flows.
This step allows us to construct a \textit{residual network}, where edges
represent transactions that exceed the levels predicted by gravity forces.

Second, we apply a weighted stochastic blockmodel (SBM)
\citep{peixoto2018PhysRevE} to the residual network to uncover its latent
community structure.
Unlike conventional community detection methods that primarily identify
densely connected groups, the SBM partitions nodes into blocks that share
statistically similar connectivity profiles, capturing the notion of
\textit{structural equivalence} in network science.
This feature is particularly suitable for production networks, where
industries and regions may occupy similar functional positions within
supply chains even if they are not directly connected.

Our analysis reveals that the Japanese production network is organized into distinct functional blocks composed of region–industry pairs that share similar macroeconomic roles. Strikingly, even after explicitly filtering out gravity forces—and thus the mechanical effect of geographic proximity—these blocks still exhibit a high degree of regional coherence. This finding indicates that multiple industries within the same region do not merely trade locally due to short distances; rather, they function collectively as deeply integrated export or import bases within the national supply chain. Crucially, alongside this regional coherence, we also uncover evidence of cross-regional integration, such as the strong structural equivalence between the Chugoku and Shikoku regions in western Japan, which act jointly as a cohesive macroeconomic zone. Furthermore, metropolitan areas such as the Kanto (greater Tokyo) and Kinki (greater Osaka) regions display a structural bifurcation between broad manufacturing bases and urban service sectors, while specific primary and infrastructure sectors (e.g., mining) span broadly across Eastern Japan. These findings suggest that effective regional coordination should not rely solely on geographic proximity. Instead, recognizing structural similarities across distant regions may help identify ``economic twins" that share common vulnerabilities and development trajectories, thereby providing a new perspective for designing wide-area regional policies.

Our study contributes to several strands of literature.
First, a large body of research has examined production networks at the
microeconomic level using firm-to-firm transaction data in Japan.
These studies provide valuable insights into the spatial organization of
supply chains and the propagation of shocks.
For example, \citet{ohnishi2010J.Econ.Interact.Coord.} identify network motifs
to characterize local interaction patterns among firms across industries.
Regarding the transmission of shocks, \citet{mizuno2014PLoSONE} highlight the
importance of microscopic linkages in shaping systemic vulnerability, while
\citet{inoue2019NatSustain} use large-scale supply chain data to simulate the
indirect economic impacts of natural disasters, demonstrating that disruptions
can propagate far beyond the directly affected regions.
Furthermore, \citet{kichikawa2019Appl.Netw.Sci.} uncover community structures
embedded in large enterprise networks and show that these communities are
closely associated with geographic boundaries.
While these micro-level studies provide important insights into firm behavior
and supply-chain resilience, our analysis complements this literature by
examining the structural organization of production networks at a higher
level of aggregation, focusing on interregional linkages across sectors.

Second, a related literature in spatial economics analyzes production networks
using interregional input--output tables.
For instance, \citet{tokui2017JapanandtheWorldEconomy} examine supply chain
disruptions following the Great East Japan Earthquake and show that complex
interregional linkages generated substantial production losses even in areas
not directly affected by the disaster.
More generally, quantitative spatial economics increasingly employs
structural general equilibrium models to conduct counterfactual analyses.
Seminally, \citet{caliendo2018Rev.Econ.Stud.} develop a framework to analyze
how regional productivity shocks propagate across the U.S.\ economy.
Following this approach, \citet{che2024JournalofRegionalScience} study the
Chinese economy to quantify the welfare effects of changes in internal trade
costs.
From a methodological perspective closer to network science,
\citet{wirkierman2021Netw.Spat.Econ.} analyze the topology of input--output
tables using Markov chains to evaluate sectoral vulnerabilities and
connectivity.
While these studies provide profound insights into aggregate welfare,
resilience, and centrality, they typically focus on shock propagation or
node-level properties rather than uncovering the latent clustering structure
of the network itself.

The remainder of this paper is organized as follows.
Section \ref{sec:empirical_strategy} introduces the interregional input--output
data and describes the empirical strategy, including the estimation of the
structural gravity model and the construction of the residual network.
Section \ref{sec:network_results} presents the stochastic blockmodel analysis
and discusses the identified community structures and their implications for
regional economic coordination.
Section \ref{sec:conclusion} concludes the paper.
Technical details on the gravity estimation and additional results are
provided in the Appendix.

\section{Empirical Strategy}
\label{sec:empirical_strategy}

\subsection{Data}
\label{subsec:data}

We use the 2005 Japanese interregional input--output (IRIO) table published by the Ministry of Economy, Trade and Industry (METI), a subnational dataset that records intermediate transactions across regions and industries.\footnote{The compilation of the official IRIO tables by METI was discontinued after the 2005 edition.}
The IRIO table records interregional transactions of intermediate inputs and
provides a comprehensive representation of production linkages across regions
and industries within Japan.

\begin{figure}[H]
\centering
\includegraphics[width=0.7\textwidth]{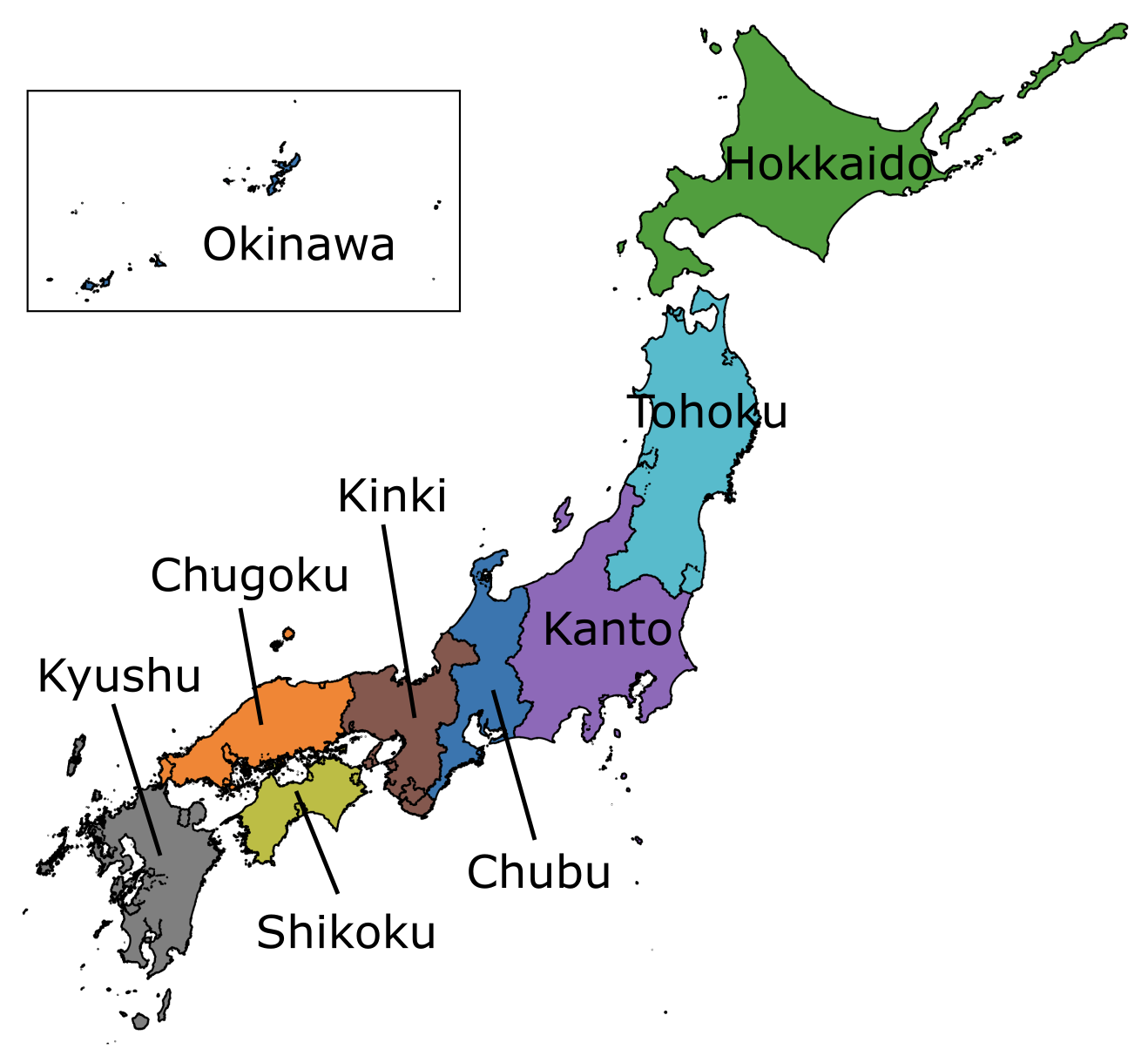}
\caption{Regional classification used in the interregional input--output table.}
\label{fig:region_map}
\end{figure}

Regions are defined as nine broad geographical blocks (Figure~\ref{fig:region_map}): Hokkaido, Tohoku, Kanto, Chubu, Kinki, Chugoku, Shikoku, Kyushu, and Okinawa. This regional aggregation follows the standard classification used in the IRIO table and reflects major economic and geographical divisions of the Japanese economy.

Turning to the sectoral dimension, we aggregate industries into 12 sectors based on the official broad classification (Table~\ref{tab:sector}).
This aggregation balances sectoral detail with empirical tractability, allowing us to clearly interpret the macroeconomic roles of the resulting blocks.

\begin{table}[H]
\centering
\caption{Sectoral classification used in the interregional input--output table.}
\label{tab:sector}
\begin{tabular}{l}
\hline
\textbf{Industry} \\
\hline
Agriculture, forestry, and fisheries \\
Mining \\
Food products \\
Metal products \\
Machinery \\
Other manufacturing \\
Construction \\
Utilities \\
Commerce and transportation \\
Finance, insurance, and real estate \\
Information and communications \\
Services \\
\hline
\end{tabular}
\end{table}

The dataset is organized at the region--industry level, where each observation corresponds to the nominal value of intermediate inputs supplied from an origin region--industry pair to a destination region--industry pair. Final demand components are excluded to focus exclusively on production linkages within the intermediate goods network. Notably, zero-valued flows are retained in the sample, which motivates the use of Poisson pseudo-maximum likelihood estimation in our empirical analysis. This data structure naturally lends itself to a network representation, with nodes corresponding to region--industry pairs and directed edges representing intermediate input flows.

\subsection{Empirical Specification}
\label{subsec:ppml}
Our empirical specification is guided by a standard quantitative trade model
with intermediate inputs in the spirit of \citet{caliendo2018Rev.Econ.Stud.};
a concise derivation is provided in Appendix~\ref{subsec:model_appendix}. We index regions by $k,\ell \in \{1,\dots,K\}$ and industries by $i,j \in \{1,\dots,I\}$.
In such models, the value of intermediate inputs $X_{ki,\ell j}$ from an origin region--industry
$(k,i)$ to a destination region--industry $(\ell,j)$ can be written as
\begin{equation}
X_{ki,\ell j}
=
\gamma_{\ell j,i}\,\pi_{\ell,ki}\,X_{\ell j},
\label{eq:theory_flow}
\end{equation}
where $\gamma_{\ell j,i}$ denotes the input share of sector $i$ in sector $j$
production, $X_{\ell j}$ is total expenditure (or gross output) of destination
$(\ell,j)$, and  $\pi_{\ell,ki}$ is the expenditure share of destination region $\ell$ on industry $i$ inputs sourced from origin region $k$.

As outlined in Appendix~\ref{subsec:model_appendix}, the theoretical framework incorporates trade frictions and multilateral resistance terms, yielding a structural gravity equation for intermediate trade. Guided by this theoretical structure, we estimate the conditional mean of
$X_{ki,\ell j}$ using Poisson pseudo-maximum likelihood (PPML).
The estimating equation is given by
\begin{equation}
\mathbb{E}\!\left[ X_{ki,\ell j} \mid \cdot \right]
=
\exp \!\left(
\alpha_{k i}
+ \delta_{\ell i}
+ \beta_i \log \tilde d_{k\ell}
+ \gamma_i Z_{k\ell}
\right)
\cdot X_{\ell ji},
\label{eq:ppml_baseline}
\end{equation}
where $X_{\ell ji} \equiv \sum_{k} X_{ki,\ell j}$ is included as an offset term.
The fixed effects $\alpha_{k i}$ and $\delta_{\ell i}$ absorb origin-side and
destination-side heterogeneity associated with origin industry $i$. The term $\tilde{d}_{k\ell}$ captures the effect of geographical distance on trade flows and is defined as $\tilde d_{k\ell} \equiv  d_{k\ell} / d_{\min}$, where $d_{k\ell}$ denotes the geodesic distance between the centroids of regions $k$ and $\ell$,\footnote{For intra-regional flows ($k=\ell$), distance is approximated using a geometric proxy based on regional area. Specifically, each region is mapped to a circle with the same area, yielding a radius $r_k = \sqrt{A_k/\pi}$, where $A_k$ denotes the area of region $k$. Intra-regional distance is then set to $d_{kk} = \frac{2}{3} r_k$, which corresponds to the expected distance between two randomly drawn points within a circle.} and $d_{\min}$ is the minimum observed distance used for scale normalization.
Distance elasticities $\beta_i$ are allowed to vary across origin industries
$i$. Furthermore, $Z_{k\ell}$ is a dummy variable indicating interregional flows involving Okinawa. Given its geographical remoteness from mainland Japan, we include this control to capture Okinawa's unique institutional and logistical frictions, thereby ensuring that these specific trade costs are not misattributed to general distance effects.

Importantly, we intentionally do not include a destination-industry fixed effect in this specification. Because the observed total expenditure $X_{\ell ji}$ by destination region $\ell$ and industry $j$ on origin industry $i$ is explicitly included as an offset term, destination-industry-specific heterogeneity is already fully absorbed. The allocation of this total demand across origin regions $k$ depends only on origin-region by origin-industry characteristics ($\alpha_{ki}$), bilateral trade frictions, and destination-region by origin-industry conditions ($\delta_{\ell i}$), making an additional destination-industry fixed effect redundant.\footnote{In the terminology of the structural gravity literature \citep{eaton2002Econometrica, anderson2003Am.Econ.Rev., caliendo2018Rev.Econ.Stud.}, outward multilateral resistance and supply-side fundamentals are absorbed by the origin fixed effects, while inward multilateral resistance and demand-side fundamentals are captured by the destination fixed effects and the observed expenditure offset, respectively. See Appendix~\ref{subsec:model_appendix} for details.}

Table \ref{tab:ppml_baseline} reports the full PPML estimation results of the gravity model. The estimated distance elasticities ($\beta_i$) are negative and highly significant across all origin industries, ranging from $-5.464$ to $-1.161$, confirming that physical distance strongly impedes interregional input-output flows. Furthermore, the coefficients for the Okinawa indicator ($\gamma_i$) are positive and significant for almost all industries (with the exception of a negative coefficient for machinery and a non-significant result for other manufacturing). The significance of these coefficients suggests that the indicator captures distinct institutional and logistical patterns associated with this remote region. In addition, their positive signs indicate that the dummy variable mitigates an excessive distance penalty; relying solely on large straight-line distances to Okinawa might otherwise overestimate the actual trade frictions, as well-developed air and maritime transport networks make the island more accessible than its geographic remoteness implies.

\subsection{Residual Network Construction}
\label{subsec:residual_network}

To identify economically meaningful connections beyond standard gravity forces,
we construct a residual network based on deviations from the estimated gravity
model. In our framework, the gravity model is not intended to provide a fully
structural causal representation of interregional input linkages.
Rather, in line with standard methodologies in network science that emphasize the use of spatial null models \citep[e.g., ][]{expert2011ProcNatlAcadSciUA}, it serves as a disciplined benchmark that captures first-order determinants of trade flows, such as geographical distance and economic scale. In this sense, the residual network highlights
systematic deviations from the gravity benchmark, which may reflect
latent economic linkages as well as other factors not captured by the
parsimonious specification.

Let $\hat{\mu}_{ki,\ell j}$ be the predicted intermediate input flow from origin
$(k,i)$ to destination $(\ell,j)$
from the PPML gravity model described above. We define the residual weight as

\[
\omega_{ki,\ell j} =
\log (X_{ki,\ell j} + \varepsilon)
-
\log \hat{\mu}_{ki,\ell j},
\]
where $\varepsilon$ is a small constant used to handle zero observations.\footnote{
The log-residual representation measures proportional deviations from the
predicted flow. Alternative normalizations such as Pearson residuals scale
deviations by $\sqrt{\hat{\mu}_{ki,\ell j}}$, reflecting the
Poisson-type variance scaling commonly used in generalized
linear models. While such scaling is natural for goodness-of-fit
diagnostics, it tends to assign disproportionately large magnitudes to
high-volume flows when the data exhibit substantial overdispersion. Because our
objective is to identify excess linkages in proportional terms rather than
absolute statistical outliers, we adopt the log-residual representation.} In our empirical implementation we set $\varepsilon = 10^{-30}$, which ensures
numerical stability while leaving the magnitude of non-zero flows essentially
unchanged.

Following the standard approach in network science of extracting a backbone composed of excess linkages \citep[e.g.,][]{serrano2009Proc.Natl.Acad.Sci.U.S.A.,fagiolo2010JEconInteractCoord}, we construct the residual network by retaining only the positive residuals ($\omega_{ki,\ell j} > 0$). These positive residuals indicate region--industry pairs whose observed intermediate input flows exceed what is predicted by geographical distance and fixed effects. We interpret such residuals as evidence of latent economic proximity, capturing patterns of production linkages that cannot be explained solely by standard gravity forces. More generally, the resulting residual network represents systematic deviations from the gravity benchmark, which may reflect omitted economic linkages as well as remaining model misspecification.

\section{Network Clustering Results}
\label{sec:network_results}

\subsection{Block Representation of the Residual Network}
To uncover the latent functional structure of the residual network, we employ a weighted stochastic blockmodel (SBM) \citep{peixoto2018PhysRevE}.
Unlike traditional community detection methods (such as modularity maximization or Infomap) that search for densely connected local cliques, the SBM is a generative model designed to cluster nodes based on \textit{structural equivalence}. This means it groups region--industry pairs that share similar probability distributions of connections to the rest of the network, regardless of whether they interact directly.
From an economic perspective, this concept perfectly captures the notion of shared macroeconomic roles. To illustrate, consider industries in geographically peripheral regions: while they may exhibit sparse direct trade among themselves, they might share a remarkably similar pattern of supply-chain dependence on specific metropolitan hubs. By focusing on these shared topological roles rather than the sheer density of direct linkages, the SBM successfully groups such nodes into the same functional community.

This conceptual alignment is particularly crucial for our analysis, as the traditional assumption of assortativity---where internal edges must be denser than external ones---is fundamentally mismatched with the nature of our gravity-filtered network.
By filtering out gravity effects, the edges with the largest residual weights tend not to be short-distance, localized transactions (e.g., those centered around Tokyo), but rather long-distance interactions that substantially exceed gravity predictions (e.g., interactions between Hokkaido and Kyushu).
Because traditional heuristics rely on internal density, these massive long-distance residual links act as heavy inter-community bridges, which causes methods like modularity or Infomap to fail entirely, typically collapsing all nodes into a single global community.
In contrast, the SBM's flexibility to identify disassortative or core--periphery structures allows it to endogenously uncover latent blocks even when the strongest economic ties traverse vast geographic distances, a structural feature highly characteristic of our residual production network.

To formalize this conceptual framework, we adopt a stochastic block representation
for the directed, weighted residual network constructed in Section~\ref{subsec:residual_network}.
This representation partitions nodes into a finite number of groups such that the
distribution of edge weights depends only on the group membership of the origin
and destination nodes.
Furthermore, to account for heterogeneity in node-level connectivity, we employ a
variant of the degree-corrected stochastic blockmodel \citep{karrer2011PhysRevEStatNonlinSoftMatterPhys}
adapted for weighted networks.

Let $b$ denote the vector of block assignments for all nodes, where $b_u$ represents the specific block to which node $u$ belongs. Conditional on these block assignments, the expected edge intensity $A_{uv}$ (which directly corresponds to the positive residual $\omega_{uv}$) from node $u$ to node $v$ is given by
\begin{equation}
\mathbb{E}[A_{uv}\mid b]=\theta^{\mathrm{out}}_u\theta^{\mathrm{in}}_v\Omega_{b_u b_v},
\end{equation}
where $\theta^{\mathrm{out}}_u$ and $\theta^{\mathrm{in}}_v$ capture the expected
out- and in-strengths (i.e., weighted degrees) of the respective nodes, and $\Omega_{rs}$
governs the baseline intensity of residual linkages between blocks $r$ and $s$.
This degree-correction mechanism ensures that
the clustering is driven purely by structural patterns of linkages---who trades with
whom---rather than simply by remaining differences in total trade volumes across
region--industry pairs.

Building on this formulation, the continuous, strictly positive edge
weights are modeled as random variables governed by their block assignments.
Given that our residual edge weights typically exhibit a right-skewed distribution,
we model these weights using an exponential distribution \citep{peixoto2018PhysRevE}.
Specifically, for a directed edge $(u\to v)$ with block assignments $(b_u,b_v)=(r,s)$,
the edge weight $A_{uv}$ follows an exponential distribution with a rate parameter $\lambda_{rs}$ specific to each pair of blocks:
\begin{equation}
p(A_{uv}|r,s)=\lambda_{rs}\exp(-\lambda_{rs}A_{uv}). 
\end{equation}
To estimate the model and determine the optimal number of blocks endogenously, we
rely on the nonparametric Bayesian inference framework developed by
\citet{peixoto2017PhysRevE, peixoto2018PhysRevE}.
Inference is performed by minimizing the description length, an information-theoretic
criterion that naturally balances goodness-of-fit with model complexity to prevent
overfitting \citep{peixoto2014Phys.Rev.X}.\footnote{To avoid local minima, we perform 300 independent runs of the optimization algorithm and select the partition that yields the minimum description length.}

\subsection{Interpretation of Clusters}
\label{subsec:interpretation}

\begin{figure}[H]
\centering
\includegraphics[scale=.5]{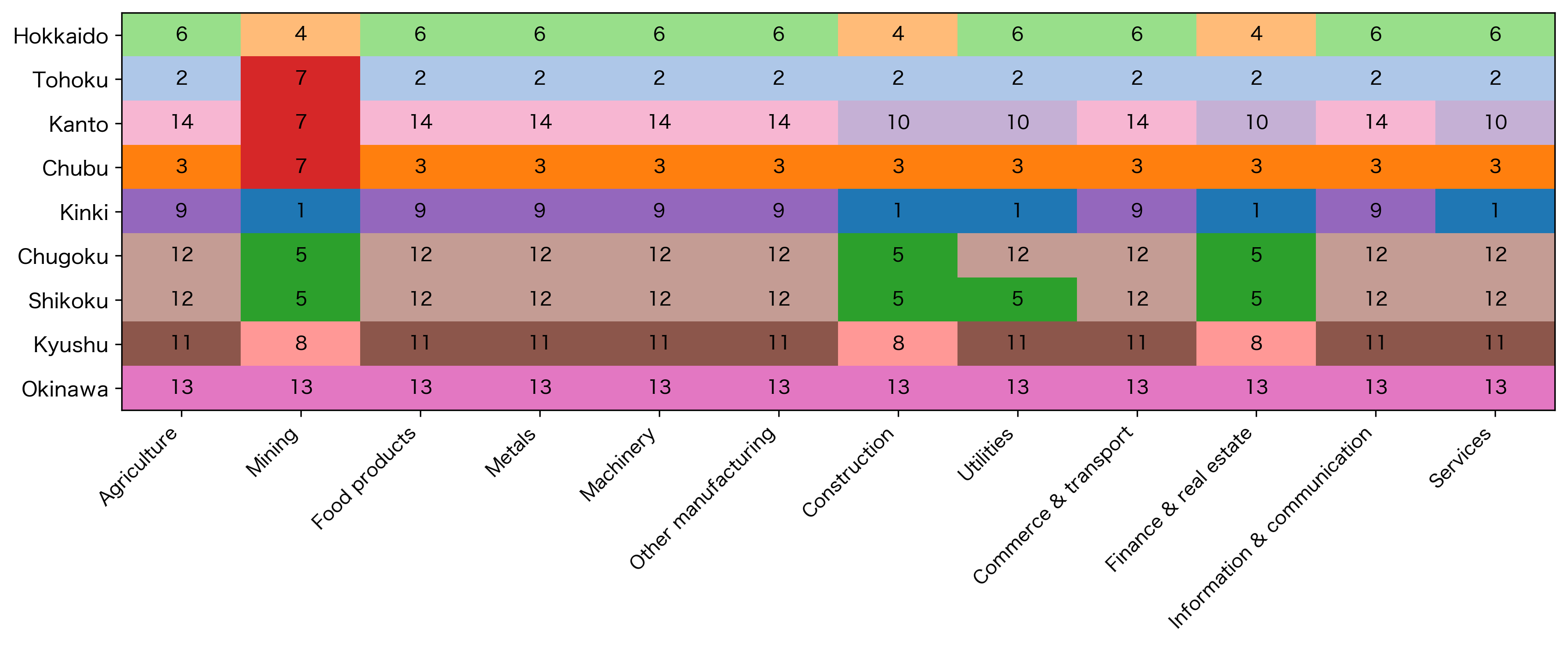}
\caption{
Cluster assignments of region--industry pairs based on the residual input--output network.
Rows correspond to regions and columns correspond to industries.
Each cell reports the cluster to which a given region--industry pair belongs, with both colors and numerical labels indicating cluster membership; the numbering of clusters is arbitrary and has no ordinal meaning.
}
\label{fig:cluster_table}
\end{figure}

Figure~\ref{fig:cluster_table} presents the clustering results in a matrix form, where rows correspond to regions and columns correspond to industries. 
By construction, these clusters summarize patterns of residual production linkages after controlling for geographical distance, economic size, and origin/destination fixed effects. 
To correctly interpret these results, it is crucial to recall that the Stochastic Blockmodel (SBM) groups nodes based on \textit{structural equivalence} rather than mere internal density. 
Therefore, region--industry pairs assigned to the same cluster are those that share a highly similar portfolio of trade connections with the rest of the national network---acting as a cohesive ``functional block'' with common macroeconomic roles.

A salient feature of Figure~\ref{fig:cluster_table} is that clusters are predominantly formed along regional lines rather than purely sectoral ones. 
Since the gravity model has already filtered out the trivial effect of spatial proximity, this regional coherence is not simply a reflection of dense localized trade. 
Instead, it implies that multiple industries within the same region exhibit an identical pattern of dependency on external markets and supply chains. 
In other words, regional economies in Japan often function as highly integrated export/import bases, interacting with national hubs collectively rather than as isolated sectoral entities.

Despite this overall regional coherence, detailed structural differentiations emerge within and across certain regions. 
In metropolitan areas such as Kanto and Kinki, there is a pronounced bifurcation between the broader manufacturing base and specific non-tradable or urban-oriented sectors. 
For instance, Kanto's construction, utilities, finance, and service sectors form a distinct cluster (Cluster 10) completely separate from its manufacturing, agriculture, and information sectors (Cluster 14). 
A parallel structural bifurcation is observed in Kinki (Clusters 9 and 1). 
This suggests that in multi-core metropolitan economies, essential urban infrastructure and service sectors operate with a fundamentally different interregional trade portfolio compared to the local production and export-oriented sectors.

Interestingly, the clustering reveals compelling evidence of cross-regional integration, most notably between the Chugoku and Shikoku regions. 
Unlike other regions that possess unique local cluster IDs, Chugoku and Shikoku share almost identical functional blocks (Clusters 12 and 5). 
This structural equivalence indicates that these two geographically adjacent regions act as a single, cohesive macroeconomic zone in the national network, jointly sharing trade dependencies and supply chain roles. 
Another notable cross-regional link is found in the mining sector across Eastern Japan, where Tohoku, Kanto, and Chubu all share the exact same assignment (Cluster 7), highlighting a shared structural role in the national resource supply network.

Conversely, peripheral regions display varying degrees of sectoral differentiation. 
In Okinawa, all industries are strictly assigned to a single, region-wide cluster (Cluster 13). 
Viewed through the lens of structural equivalence, this indicates that the entire island economy shares a uniform structure of dependence on the mainland, likely driven by specific institutional conditions, public expenditure, and geographical constraints. 
In Hokkaido and Kyushu, while the primary base remains integrated, specific sectors such as mining, construction, and finance separate into sub-clusters, reflecting a localized division of roles between general production and regional infrastructure maintenance.

Taken together, the SBM reveals that residual interregional input linkages are organized around functionally equivalent regional blocks. 
These groupings capture latent, macro-level economic interdependencies—such as the tight integration of Chugoku and Shikoku or the urban bifurcation in Kanto and Kinki—defining how local economies collectively position themselves within the complex web of the Japanese production network. 
Building on this structural insight, the next section examines the profound implications of these findings for wide-area regional coordination.

\subsection{Implications for Regional Coordination}
\label{subsec:implications}

The clustering results provide a fundamentally new, data-driven perspective on regional coordination. 
Conventional policies, such as the Network-based Core City Region Initiative, largely assume that coordination should be vertically integrated along administrative borders or physical proximity. 
However, by analyzing the network through the lens of structural equivalence, our findings reveal that the underlying architecture of the Japanese economy is organized into distinct functional blocks sharing common macroeconomic roles. 
This shift in perspective---from spatial closeness to network-based functional equivalence---offers important strategic implications for wide-area economic policies, particularly in terms of operational efficiency, risk management, and cross-regional knowledge sharing.

The identification of these functional blocks provides a network-based perspective on regional coordination, highlighting both shared opportunities and common vulnerabilities. Because industries within the same structural block share similar portfolios of external dependencies, they face common structural bottlenecks. Recognizing these similarities could facilitate broader horizontal coordination---such as joint logistics or supply-chain negotiations---allowing region--industry pairs to pool resources rather than competing individually. At the same time, this structural equivalence inherently implies shared exposure to systemic risks; a localized shock to a common external hub could simultaneously affect the entire cluster. This vulnerability is particularly evident for regions like Okinawa, where industries are heavily concentrated in a single functional block, implying a monolithic exposure to external disruptions. Consequently, wide-area policy frameworks might benefit from targeting structurally equivalent regions collectively, not only to enhance operational efficiency but also to collaboratively build supply-chain resilience against aggregate shocks.

Furthermore, the presence of cross-regional clusters---such as the shared mining roles across Eastern Japan or the functional integration between Chugoku and Shikoku---highlights the potential for coordination among \textit{economic twins}.
Region--industry pairs that fall into the same cluster but are physically separated share a similar ``economic DNA'' within the national network.
Because they likely face parallel structural challenges, such as demographic constraints or over-reliance on specific supply chains, recognizing these non-spatial linkages could help policymakers facilitate knowledge spillovers and share best practices between structurally equivalent municipalities.
Such an approach offers a valuable complement to traditional administrative frameworks, which often exhibit a neighbor-only bias.

\section{Conclusion}
\label{sec:conclusion}
In this paper, we have provided a data-driven perspective on the architecture of regional economies and wide-area coordination, using the Japanese production network as an empirical setting.
Recognizing that conventional regional policies often implicitly assume that
economic linkages are dictated by geographic proximity, we set out to uncover
the actual, latent interdependencies embedded in the nationwide production network.
To isolate these non-trivial connections, we first estimated a structural gravity
model using highly disaggregated interregional input--output data, effectively
filtering out the pervasive effects of spatial frictions and economic size.
We then analyzed the resulting residual network through the lens of a weighted
SBM.

Our empirical findings reveal that the Japanese production network is endogenously organized into distinct ``functional blocks'' that exhibit a high degree of regional coherence. Strikingly, even after explicitly filtering out gravity effects, multiple industries within the same region tend to cluster together based on their shared macroeconomic roles. This suggests that these regions do not merely trade locally due to geographic proximity, but rather function collectively as integrated export or import bases within the national supply chain. Beyond this baseline regional coherence, our analysis highlights the diverse and complex roles of local economies. For instance, we find clear evidence of cross-regional integration, such as the structural equivalence between the Chugoku and Shikoku regions, which jointly function as a cohesive macroeconomic zone. Furthermore, the results capture a structural bifurcation in metropolitan areas like Kanto and Kinki, alongside the broad geographic span of specific primary and infrastructure sectors across Eastern Japan.

Despite these insights, several avenues for future research remain.
First, while our analysis relies on broad regional classifications, applying
our framework to more granular spatial and sectoral data would allow for a more high-resolution
mapping of structural equivalence.\footnote{For instance, future research could utilize the inter-prefectural input--output tables provided by the Research Institute of Economy, Trade and Industry (RIETI) in the case of Japan.} 
Such an extension could reveal whether the functional blocks identified at the
macro-regional level hold uniformly, or if intra-regional bifurcations exist
at the prefectural level.
Second, introducing a temporal dimension is a crucial next step. 
Applying dynamic community detection to historical input--output tables would
elucidate how these functionally equivalent blocks evolve in response to macro
shocks, such as natural disasters or globalization.
Third, cross-country comparisons using international spatial data are needed
to determine whether the emergence of structurally equivalent regional blocks
is a universal topological feature of modern spatial economies or a consequence
of Japan’s unique geographic and institutional characteristics.
Finally, integrating these endogenously identified functional blocks into
quantitative spatial economic models (QSMs) would offer a promising framework
for simulating the counterfactual effects of wide-area coordination policies
and shock propagations.

\appendix
\section{Appendix}
\label{app:model}

\setcounter{table}{0}
\renewcommand{\thetable}{A.\arabic{table}}
\setcounter{figure}{0}
\renewcommand{\thefigure}{A.\arabic{figure}}

\subsection{Model-Based Motivation}
\label{subsec:model_appendix}
This appendix provides a concise theoretical motivation for the empirical
specification used in the main text.
The exposition follows the logic of standard quantitative trade models with
intermediate inputs, without presenting the full complexity of the model.

\subsubsection{Intermediate Input Flows}

Consider an economy with multiple regions and industries.
Let $(k,i)$ index an origin region and industry, and $(\ell,j)$ index a
destination region and industry.

Following the framework of \citet{caliendo2018Rev.Econ.Stud.}, intermediate input flows are determined by two distinct mechanisms: technological input requirements and spatial sourcing decisions. First, due to a Cobb-Douglas production technology, destination $(\ell, j)$ allocates a constant share $\gamma_{\ell j,i}$ of its total expenditure (or gross output), $X_{\ell j}$, to inputs from industry $i$. Second, under iceberg trade costs and a Fréchet productivity structure \citep{eaton2002Econometrica}, the destination sources these inputs from various origin regions $k$ with a probabilistic sourcing share $\pi_{\ell,k i}$ that takes a gravity form:
\begin{equation}
\pi_{\ell,k i} = \frac{(c_{k i}\tau_{k\ell})^{-\theta_i}(T_{k i})^{\theta_i\sum_j \gamma_{k j, i}}}{\sum_{k'} (c_{k'i}\tau_{k'\ell})^{-\theta_i}(T_{k'i})^{\theta_i\sum_j \gamma_{k'j, i}}}, \label{eq:app_pi}
\end{equation}
where $T_{ki}$ represents the fundamental productivity level governing the Fr\'{e}chet distribution, $c_{ki}$ is the unit cost of the composite input bundle (comprising labor and intermediate goods), $\tau_{k\ell}$ denotes the iceberg trade costs from $k$ to $\ell$, and $\theta_i$ is the trade elasticity for sector $i$, which inversely measures the dispersion of productivity within the sector. The denominator corresponds to a destination-region and origin-industry specific multilateral resistance term.

Combining these technological and spatial components, the value of intermediate input flows from $(k, i)$ to $(\ell, j)$ can be seamlessly written as:
\begin{equation}
X_{ki,\ell j} = \gamma_{\ell j,i} \pi_{\ell,ki} X_{\ell j}. \label{eq:app_flow}
\end{equation}

\subsubsection{Mapping to the Empirical Specification}

Taking logs of \eqref{eq:app_flow} and using \eqref{eq:app_pi}, the value of intermediate input flow from $(k, i)$ to $(\ell, j)$ can be expressed in a log-linear form. This structure directly motivates the empirical specification in equation \eqref{eq:ppml_baseline}.  To make the correspondence between the theoretical model and the empirical estimating equation clearer, we outline the mapping of the fixed effects, the offset term, and the distance term.

Specifically, the origin region-industry fixed effect, $\alpha_{ki}$, absorbs the logarithmic transformation of $(c_{ki})^{-\theta_i}(T_{ki})^{\theta_i\sum_j \gamma_{kj, i}}$. Similarly, the destination-region by origin-industry fixed effect, $\delta_{\ell i}$, absorbs the multilateral resistance term that explicitly corresponds to the logarithm of the denominator of \eqref{eq:app_pi}. 

Furthermore, the term $\gamma_{\ell j, i}X_{\ell j}$ in the theoretical model represents the scale of intermediate input demand, that is, the total expenditure allocated by destination $(\ell, j)$ to inputs from industry $i$. In our empirical specification, this maps precisely to the offset variable $X_{\ell ji}$. By including $X_{\ell ji}$ as an offset (a predictor with its coefficient constrained to 1), the model explicitly accounts for the total expenditure capacity of the destination. 

Finally, the bilateral trade cost $\tau_{k\ell}$ in the theoretical model is proxied by geographical distance and other bilateral variables in the empirical model. Consequently, the estimated parameters, such as the distance elasticity $\beta_i$, capture the pure effect of spatial frictions on the relative distribution of these intermediate input flows, having fully controlled for structural differences in expenditure and production sizes.

\subsection{Full Estimation Results of the Gravity Model}
\label{app:estimation_results}

This appendix provides the full estimation results of the gravity model discussed in Section \ref{subsec:ppml}. Table \ref{tab:ppml_baseline} reports the detailed coefficients and standard errors for all origin industries.
\begin{table}[H]
\centering
\caption{PPML Estimation Results of the Gravity Model}
\label{tab:ppml_baseline}
\begin{tabular}{lcc}
\toprule
\textbf{Dependent Variable:} & \multicolumn{2}{c}{$X_{ki,\ell j}$} \\
\textbf{Estimator:} & \multicolumn{2}{c}{\textbf{PPML}} \\
\midrule
\textbf{Origin Industry ($i$)} & $\log \text{Distance}$ ($\beta_i$) & \text{Okinawa Dummy} ($\gamma_i$) \\
\midrule
Agriculture  & $-1.725^{***}$ & $1.822^{***}$ \\
            & $(0.101)$      & $(0.312)$     \\
Mining  & $-4.432^{***}$ & $7.350^{***}$ \\
            & $(0.253)$      & $(0.788)$     \\
Food products  & $-1.422^{***}$ & $0.992^{***}$ \\
            & $(0.067)$      & $(0.210)$     \\
Metals  & $-1.476^{***}$ & $0.482^{***}$ \\
            & $(0.025)$      & $(0.116)$     \\
Machinery  & $-1.161^{***}$ & $-0.801^{***}$ \\
            & $(0.076)$      & $(0.298)$     \\
Other manufacturing  & $-1.274^{***}$ & $0.093$  \\
            & $(0.046)$      & $(0.128)$     \\
Construction  & $-5.464^{***}$ & $7.305^{***}$ \\
            & $(0.475)$      & $(0.994)$     \\
Utilities  & $-3.299^{***}$ & $1.497^{**}$ \\
            & $(0.235)$      & $(0.609)$     \\
Commerce \& transport  & $-1.196^{***}$ & $0.456^{***}$ \\
            & $(0.036)$      & $(0.136)$     \\
Finance \& real estate & $-4.392^{***}$ & $5.359^{***}$ \\
            & $(0.489)$      & $(0.975)$     \\
Information \& communication & $-2.167^{***}$ & $2.241^{***}$ \\
            & $(0.241)$      & $(0.619)$     \\
Services & $-2.655^{***}$ & $3.585^{***}$ \\
            & $(0.269)$      & $(0.670)$     \\
\midrule
\textbf{Fixed Effects} & & \\
Origin $\times$ Origin-Industry ($\alpha_{ki}$) & Yes & Yes \\
Dest $\times$ Origin-Industry ($\delta_{\ell i}$) & Yes & Yes \\
\midrule
\textbf{Summary Statistics} & & \\
Observations       & \multicolumn{2}{c}{10,719} \\
Pseudo $R^2$       & \multicolumn{2}{c}{0.983} \\
\bottomrule
\multicolumn{3}{p{12cm}}{\footnotesize \textit{Notes:} Standard errors clustered at the origin region-industry level are reported in parentheses. Significance levels: $^{*} p < 0.10$, $^{**} p < 0.05$, $^{***} p < 0.01$. The dependent variable is the value of intermediate input flow. Total destination-side expenditures are included as an offset term.}\\
\end{tabular}
\end{table}

\bibliographystyle{abbrvnat}
\bibliography{epj_references}
\end{document}